\documentclass[
prl,letterpaper,twocolumn,superscriptaddress,amsmath,amssymb,floatfix
]{revtex4-2}

\usepackage{xcolor}  % This is used to mark the change by the revision
\usepackage{graphicx}% Include figure files
\usepackage{dcolumn}% Align table columns on decimal point
\usepackage{bm}% bold math
\newcommand{\ket}[1]{\ensuremath{\left|#1\right\rangle}}
\definecolor{black}{rgb}{0.,0.,0}
\definecolor{orange}{rgb}{0.94,0.40,0}
\definecolor{darkgreen}{rgb}{0,0.50,0}

\begin{document}

%\title{Cancelling microwave crosstalk with fixed-frequency qubits}
\title{Cancelling microwave crosstalk with fixed-frequency qubits}

%%%%%%%%%%%%%%%%%%%%%%%%%%%%%%%%%%%%%%%%%%%%%%%%%%%%%%%%%%%%%%%%%%%
% 518055

\newcommand{\SIQSE}{\affiliation{1}{Shenzhen Institute for Quantum Science and Engineering, Southern University of Science and Technology, Shenzhen, Guangdong, China}}
% 518055
\newcommand{\IQA}{\affiliation{2}{International Quantum Academy, Shenzhen, Guangdong, China}}
% 518048
\newcommand{\GDKL}{\affiliation{3}{Guangdong Provincial Key Laboratory of Quantum Science and Engineering, Southern University of Science and Technology, Shenzhen, Guangdong, China}}
\newcommand{\SUS}{\affiliation{4}{Department of Physics, Southern University of Science and Technology, Shenzhen, Guangdong, China}}
\newcommand{\NJU}{\affiliation{5}{National Laboratory of Solid State Microstructures, School of Physics,
	Nanjing University, Nanjing 210093, China}}

\author{Wuerkaixi Nuerbolati}
\affiliation{\NJU}

\author{Zhikun Han}
\affiliation{\SIQSE}\affiliation{\IQA}\affiliation{\GDKL}

\author{Ji Chu}
\affiliation{\SIQSE}\affiliation{\IQA}\affiliation{\GDKL}

\author{Yuxuan Zhou}
\affiliation{\SIQSE}\affiliation{\IQA}\affiliation{\GDKL}

\author{Xinsheng Tan}
\email{tanxs@nju.edu.cn}
\affiliation{\NJU}

\author{Yang Yu}
\affiliation{\NJU}

\author{Song Liu}
\affiliation{\SIQSE}\affiliation{\IQA}\affiliation{\GDKL}

\author{Fei Yan}
\email{yanf7@sustech.edu.cn}
\affiliation{\SIQSE}\affiliation{\IQA}\affiliation{\GDKL}

%%%%%%%%%%%%%%%%%%%%%%%%%%%%%%%%%%%%%%%%%%%%%%%%%%%%%%%%%%%%%%%%%%

\begin{abstract}
Scalable quantum information processing requires 
	that modular gate operations can be executed in parallel. 
The presence of crosstalk decreases 
	the individual addressability, 
	causing erroneous results during simultaneous operations.
For superconducting qubits which operate 
	in the microwave regime, 
	electromagnetic isolation is often limited 
	due to design constraints, 
	leading to signal crosstalk that 
	can deteriorate the quality of simultaneous gate operations.
Here, we propose and demonstrate a method based on 
	AC Stark effect for calibrating 
	the microwave signal crosstalk.
The method is suitable for processors 
	based on fixed-frequency qubits which 
	are known for high coherence and simple control.
The optimal compensation parameters can be reliably 
	identified from a well-defined interference pattern.
We implement the method on an 
	array of 7 superconducting qubits, 
	and show its effectiveness in removing the majority of 
	crosstalk errors.
\end{abstract}

\maketitle
Crosstalk is a major factor that 
	impedes the development of 
	scalable quantum information processing architectures. 
Among the miscellaneous crosstalk phenomena 
	including measurement crosstalk
	\cite{altomare2010measurement,khezri2015qubit}, 
	signal crosstalk 
	\cite{wenner2011wirebond,dai2021calibration,abrams2019methods,huang2021microwave,spring2021high}
	and quantum-mechanical crosstalk (spurious coupling)
	\cite{mundada2019suppression,patterson2019calibration,zhao2021quantum,cai2021impact}, 
	the microwave signal crosstalk is notorious 
	for superconducting qubits which usually operate at 4--8~GHz.
Various technologies have been developed to 
	improve isolation between circuit components 
	to reduce such crosstalk 
	\cite{chen2014fabrication, dunsworth2018method, rosenberg2020solidstate, wei2021quantum,mitchell2021hardware}.
As the integration level of quantum processors
	continue to increase, 
	the presence of denser wires and more crowded spectrum 
	unavoidably leads to adverse crosstalk effect.
To remove microwave signal crosstalk,
	an effective approach is to offset the 
	crosstalk signal with an out-of-phase compensation signal. 
This has been demonstrated in processors 
	with tunable qubits where one is able to apply 
	resonant driving and monitor 
	the effectiveness of compensation 
	\cite{sung2020realization}.
However, the dynamic tuning range is limited due to design and fabrication variation, and the resonance condition may not be satisfied.
Moreover, such a protocol is unsuitable 
	for architectures based on fixed-frequency qubits 
	which have recently been demonstrated 
	with high-fidelity entangling operations using simple control \cite{collodo2020implementation,xu2020high,stehlik2021tunable}.

In this work, we propose and demonstrate a method of 
	microwave crosstalk calibration that is applicable 
	to fixed-frequency qubits.
Our protocol is based on the alternative-current (AC) Stark effect \cite{schuster2005ac} or the Autler-Townes effect, by which a detuned crosstalk drive can induce a shift in the qubit frequency. 
Using a spin-echo sequence, we measure the Stark-induced phase, which shows a well-defined 
	interference pattern with respect to 
	the compensation drive parameters, 
	which can be reliably optimized with increasing sequence duration. 
The qubit frequency measured using the Ramsey fringe 
	is found independent of the crosstalk signal amplitude, 
	confirming the effectiveness in crosstalk compensation.
Finally, we perform randomized benchmarking (RB) 
	with simultaneous single-qubit gates 
	on a chain of 7 fixed-frequency qubits. 
The result shows clear improvement from crosstalk compensation
	for all qubits, with up to 90\% 
	of signal crosstalk errors removed. 

\begin{figure}[htbp]
\includegraphics[scale=0.5]{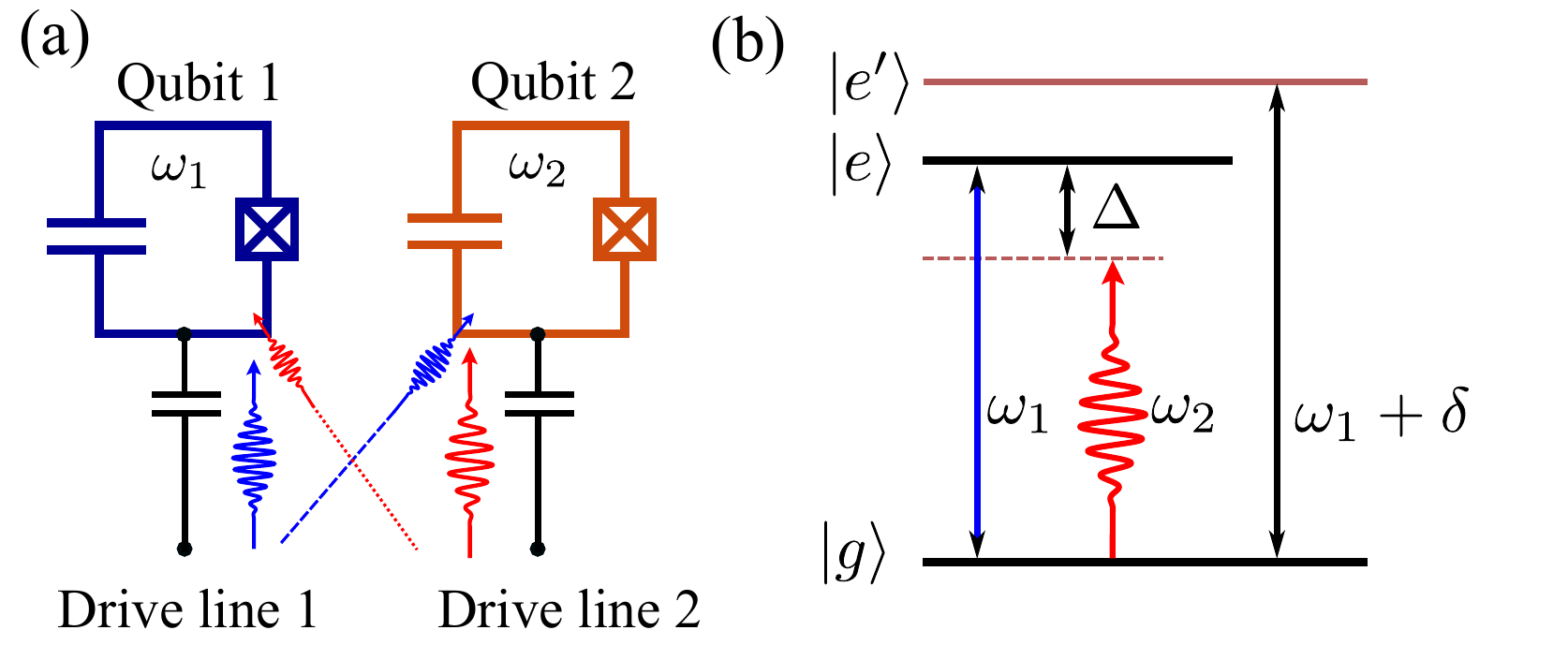}
\caption{
{ \bf(a) } 
Schematic of microwave signal crosstalk. 
Two fixed-frequency transmon qubits 
	have local drive lines sending microwave pulses 
	at their respective frequency. 
Different colors indicate different frequencies. 
A smaller signal cross-talks to the other qubit.
{ \bf(b)}
The AC Stark effect in which the excited state ($\ket{\rm e}$) of the target qubit is pushed higher to $\ket{\rm e'}$ under
	a red-detuned crosstalk drive, resulting in a positive frequency shift $\delta$.
}
\label{fig:f1} 	
\end{figure}

% the problem : 
We study the problem by first considering two fixed-frequency transmon qubits \cite{koch2007chargeinsensitive} 
	where the nominally dedicated drive lines 
	may couple to the other qubit, 
	as illustrated in Fig.~\ref{fig:f1}a. 
In general, 
	the two qubits have different frequencies, 
	$\omega_1$ and $\omega_2$ (assuming $\omega_1>\omega_2$).
Suppose that resonant pulses are 
	applied to Qubit-2 for coherent operations. 
At the same time, 
	Qubit-1 is also driven by these pulses due to crosstalk, though off-resonantly.
According to the AC Stark effect, 
	such off-resonant drive
	induces a shift in the effective frequency of Qubit-1 by 
\begin{equation}
\begin{split}
\delta(\Omega,\Delta) &= {\rm sgn}(\Delta) (\sqrt{\Omega^{2}+\Delta^{2}} - |\Delta|) \\
&\approx \frac{\Omega^2 }{2 \Delta} \;, \label{eq:acsol}
\end{split}
\end{equation}
	where $\Delta=\omega_1-\omega_2$ is the detuning between 
	the qubits and $\Omega$ is the equivalent Rabi frequency of the crosstalk drive. 
As displayed in the shown example (Fig.~\ref{fig:f1}b), the Stark shift $\delta$ is positive when the drive is red-detuned ($\Delta>0$), and negative the otherwise.
The second line of Eq.~\eqref{eq:acsol} is the approximated 
	form when $\Omega\ll|\Delta|$. 
Such a frequency shift is a main source of 
	gate error because it changes the actual operating frequency of the qubit. 

The crosstalk signal may be cancelled by interfering it with a compensation signal with the same shape, 
	frequency and amplitude but $180^\circ$ out-of-phase. 
However, because of the difference in overall attenuation 
	and electromagnetic wave delaybetween drive line paths, 
	the compensation signal has to be calibrated 
	by using the qubit as the probe.
In practice, because the wave delay between different drive 
  lines (1~ns) is much shorter than the typical gate duration (30~ns), 
  it is sufficient to calibrate the amplitude and phase only.

%Explainging how to measure crosstalk % The method
\begin{figure}[htbp]
\includegraphics[scale=1]{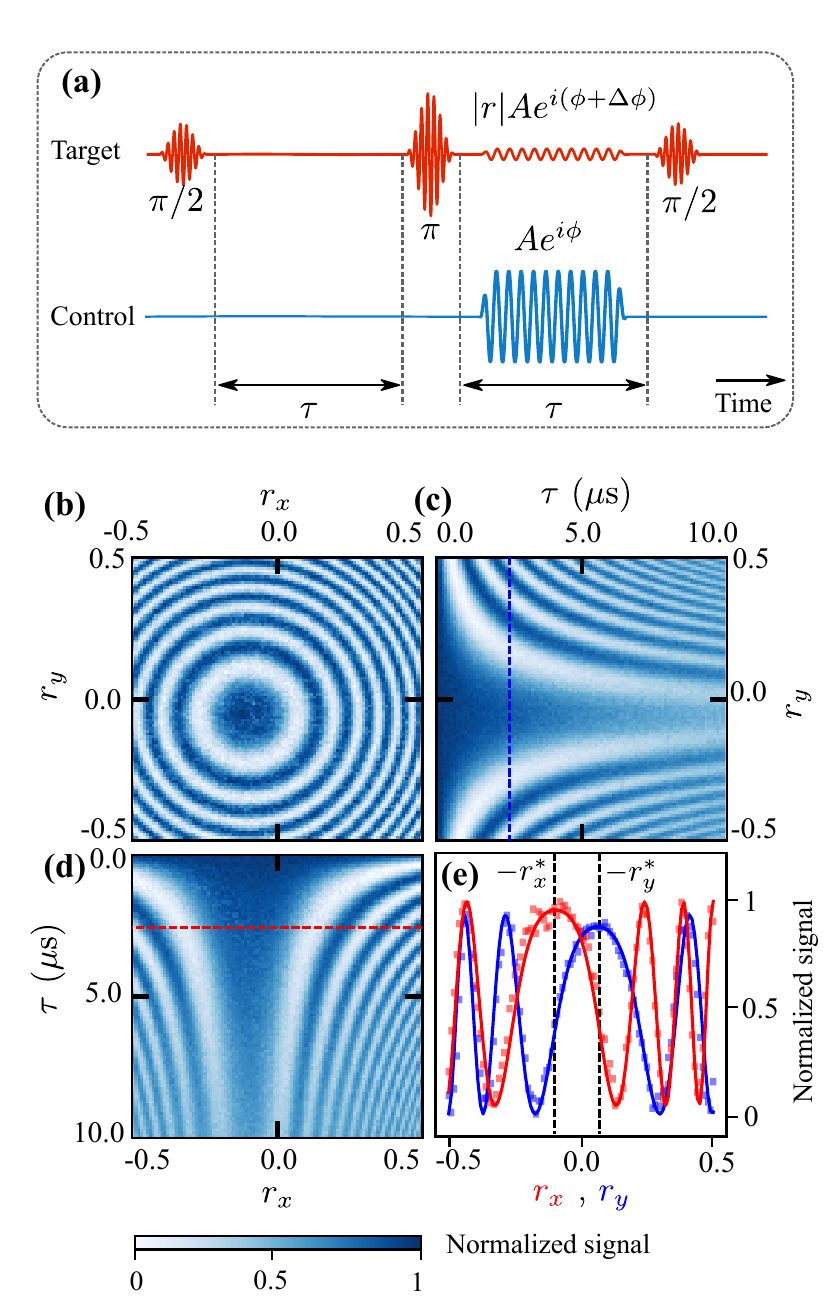}% Here is how to import EPS art
\caption{
{\bf (a)} 
Pulse sequence for calibrating the crosstalk.
	A spin-echo pulse sequence 
	(total free-evolution time: $2\tau$) 
	is applied to the target qubit.
A square-shaped resonant pulse with certain amplitude ($A$) 
	and phase ($\phi$) is applied to the control qubit during 
	the second half of free-evolution period. 
The compensation pulse with relative amplitude ($|r|$) 
	and phase ($\Delta\phi$) but at the control qubit 
	frequency is simultaneously applied to the target qubit. 
{\bf (b)} 
Measured response of the target qubit 
	in the complex plane of the compensation phasor 
	$\vec{r}=|r|\exp\{i\Delta\phi\}$, which resembles the Newton's rings.
The center of the rings 
	indicates where the crosstalk signal from 
	the control qubit (phasor $\vec{r}^*$) is completely compensated, 
	i.e., $\vec{r}+\vec{r}^*=0$.
{\bf (c,d)}
Target qubit response measured at $r_x=0$ ({\bf (c)}) 
	and $r_y=0$ ({\bf (d)}) with varying free-evolution time $\tau$,
	showing faster fringes with increased $\tau$.
{\bf (e)}
A linecut from {\bf (c)} (blue) 
	and {\bf (d)} (red) measured at $\tau = 2.5~\mu$s.
The solid lines are fit 
	to Eq.~\eqref{eq:SignalFit} for finding the optimal 
	compensation parameters $-\vec{r}^*$.
}
\label{fig:f2}
\end{figure}

We first demonstrate our method in a two-qubit subsystem 
	out of a 16-qubit processor, the design of which is similar to the device presented in Ref.~\cite{chu2021scalable}. 
The two transmon qubits have frequencies: 
	6.2497~GHz (target qubit) and 6.2718~GHz (control qubit).
 
They are separated in a qubit chain, and their coupling is negligible by design ($\sim100$~kHz). 
We apply to the target qubit the spin-echo 
	sequence (Fig.~\ref{fig:f2}a), which is known for its robustness against low-frequency noise 
	and therefore provides better coherence time 
	and signal visibility.
Given a test pulse with certain amplitude $A$ 
	and phase $\phi$ applied to the control qubit drive line, crosstalk of this drive induces a Stark shift $\delta$ and hence a non-refocused phase on the target qubit during the echo sequence. 
Our goal is to search for the optimal compensation drive (phasor $\vec{r}=|r|\exp\{i\Delta\phi\}$) applied to the target qubit, which offsets the crosstalk drive (phasor $\vec{r}^*$).
As shown in Fig.~\ref{fig:f2}b, 
	for a fixed $\tau$,	the measured response in the complex plane of the compensation phasor shows an interference pattern that resembles the Newton's rings.
The center of the rings indicates 
	the optimal working point for compensation;	
	an additional $2\pi$ phase is accumulated for each ring outward. 
	The pattern can be described by
\begin{equation}
\begin{split}
    S = \frac{1}{2} \left[ 1+\cos\left( \delta\left(|\vec{r}+\vec{r}^*|A, \Delta\right) \cdot \tau \right) \right] \,, \label{eq:SignalFit}
\end{split}
\end{equation}
where $S$ is the normalized signal.

Equation \eqref{eq:SignalFit} implies that 
	a longer free-evolution time $\tau$ reduces 
	the oscillating period in the Newton's rings (faster ripples),
	effectively increasing the sensitivity to the Stark shift. 
Such an effect is illustrated in Fig.~\ref{fig:f2}c 
	and Fig.~\ref{fig:f2}d where we fix the real ($r_x$) and imaginary ($r_y$) part of 
	the compensation phasor, 
	respectively, but vary $\tau$.
Figure \ref{fig:f2}e shows a vertical ($r_x=0$) and horizontal ($r_y=0$) linecut measured at $\tau=2.5~\mu$s,
	which are fitted to Eq.~\eqref{eq:SignalFit} 
	for extracting the optimal compensation drive parameters.
In practice, we progressively increase $\tau$ 
	during the parameter search to improve the accuracy, and we perform a gradient-based search algorithm during iterative measurements with a feedback loop to speed up the optimization process.

\begin{figure}[htbp]
\includegraphics[scale=1]{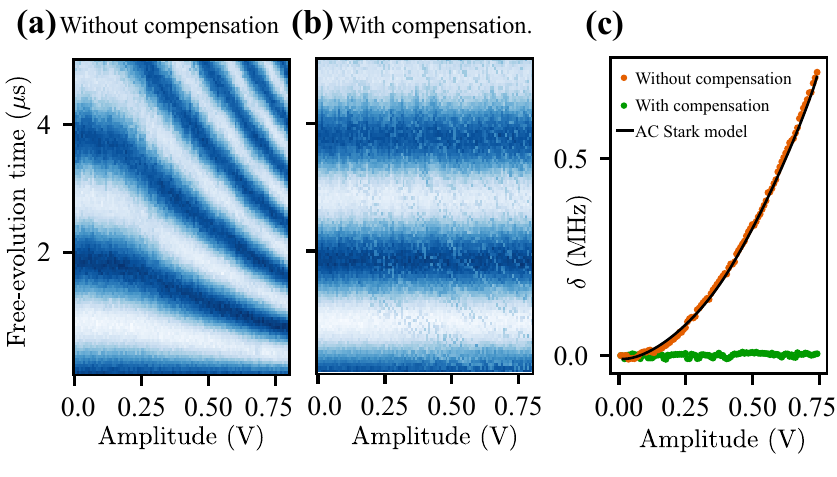}% Here is how to import EPS art
\caption{ 
Measured Ramsey fringes of the target qubit 
	at different amplitude of the control drive {\bf (a)} 
	without and {\bf (b)} with compensation.
We intentionally add a baseline 500~kHz detuning for presentation purpose.
{\bf (c)}
The extracted frequency shift in both cases 
	showing the removal of the crosstalk effect across the whole amplitude range.
The solid line is the fit to the Eq.~\eqref{eq:acsol}.
\label{fig:f3}
}
\end{figure}

Next, we verify our calibration by conducting 
	a standard Ramsey experiment.
The Ramsey fringe indicates the target-qubit frequency shift.
We vary the amplitude of the test pulse applied 
	to the control qubit and compare 
	the measured Ramsey fringes without (Fig.~\ref{fig:f3}a) 
	and with (Fig.~\ref{fig:f3}b) crosstalk compensation.
Obviously, without compensation, 
	the fringes become faster with increased drive amplitude, 
	indicating stronger AC stark effect.
With compensation, 
	the fringes becomes stabilized and unaffected 
	by the test pulse.
We note that there exists small ripples, 
	possibly as a result of imperfect compensation.
The frequency shift $\delta$ in both cases 
	are extracted and compared in Fig.~\ref{fig:f3}c. 
	Without compensation, 
	the measured result is in good agreement 
	with the AC stark effect as described in Eq.~\eqref{eq:acsol}.
The independence of the drive amplitude validates the linear response model of the crosstalk effect and assures the use of the optimization result to arbitrary waveforms.

Equation \ref{eq:SignalFit} indicates that the smallest detectable Stark shift depends on the sequence length $\tau$. Therefore, the sensitivity of our method is proportional to the coherence time of the qubit.
We may compare our method to the Rabi-based method. 
In the case of tunable qubits, the resonance condition $\Delta=0$ is an extreme case in Eq.~\ref{eq:acsol}, where the Stark shift is maximized at $\delta=\Omega$. Experimentally, this may be done with a slow enough rise and fall for adiabatically ramping up the Stark shift. In this case, the Stark shift is equivalent to the Rabi method whose signal also oscillates at the Rabi frequency $\Omega$. Both are limited by the coherence time.
However, in the case of non-zero $\Delta$ such as for fixed-frequency qubits, the visibility of the Rabi oscillations are significantly reduced to $(\Omega/\Delta)^2$ in addition to decoherence, while the method based on Stark shift is unaffected.

%%  Explaining the Experiment
\begin{figure}[htbp]
\includegraphics[scale=0.95]{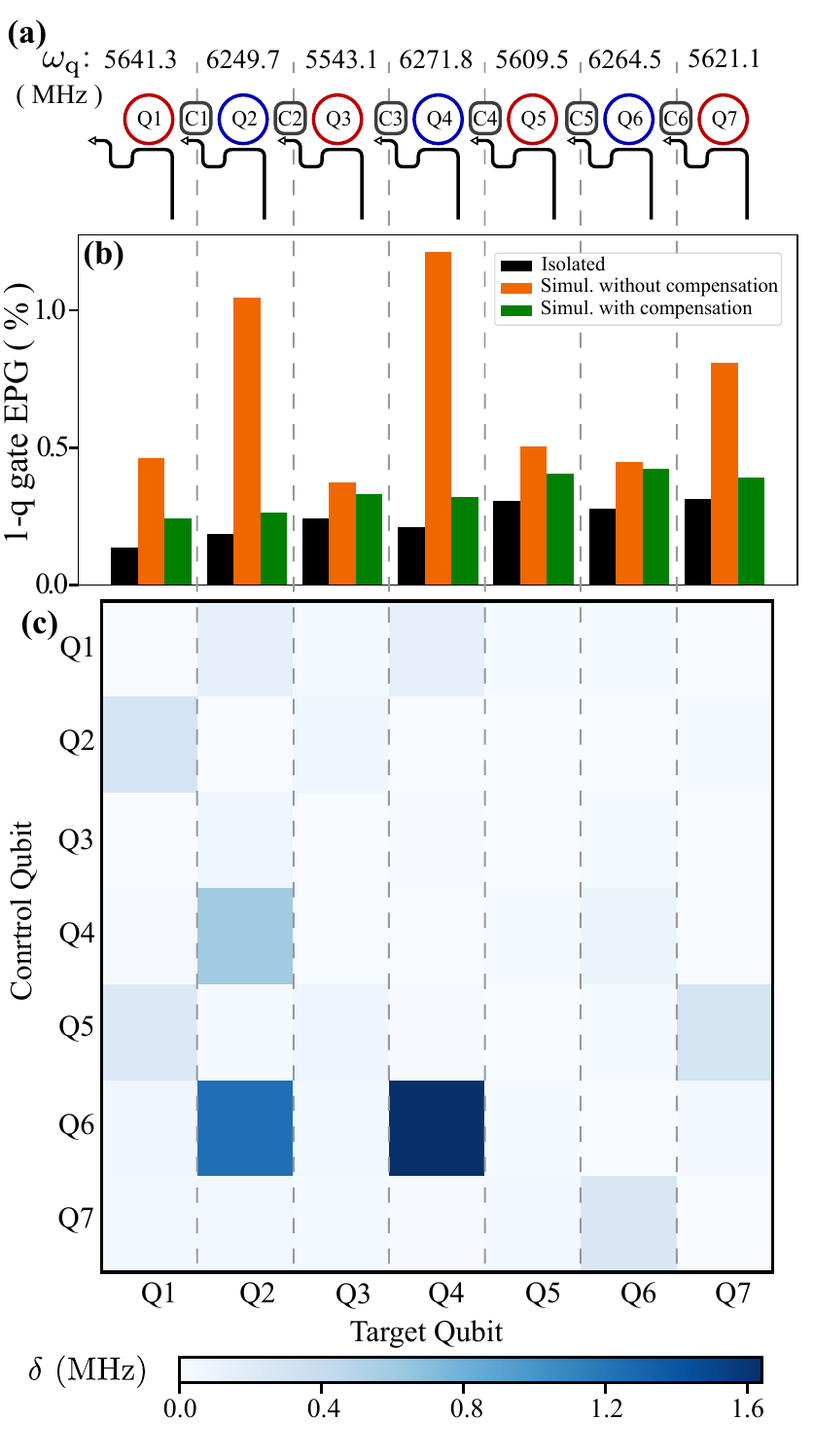}% Here is how to import EPS art
\caption{  
{\bf (a)} 
Schematic of an array of 7 fixed-frequency transmon qubits 
	interleaved with tunable couplers. 
A shared control line is used to deliver both the microwave pulse to the qubit and the slow flux pulse to the coupler.
{\bf (b)}
Error per gate (EPG) of single-qubit gates obtained by implementing RB separately (black), 
	simultaneously without compensation (orange), 
	and simultaneously with compensation (green).
See Table.~\ref{tab:t1} for listed values.
{\bf (c)}
The AC Stark shift matrix as an indicator of the crosstalk impact on the target qubit when driving the control qubit resonantly.
For example, crosstalk from Q4 to Q6 shows the strongest shift due to their relatively small detuning ($\Delta=5.5$~MHz), which is consistent with the error deterioration of Q4 in the simultaneous RB result.
}
\label{fig:f4}
\end{figure}
%%%%%%%%%%%%%%%%%%%Table%%%%%%%%%%%%%%%%%%%%%%%%

\begin{table}[htbp]
{ 
	\begin{tabular}{|c|c|c|c|c|c|c|c|}
	\hline
	RB error & Q1 & Q2 & Q3 & Q4 & Q5 & Q6 & Q7  \\ 
	\hline
	$ {\color{black} \blacksquare}$ (\%) & 0.13 & 0.19 & 0.24 & 0.21 & 0.31 & 0.28 & 0.31  \\ 
	\hline
	$ {\color{orange} \blacksquare}$ (\%) & 0.46 & 1.05 & 0.37 & 1.21 & 0.50 & 0.45 & 0.81  \\ 
	\hline
	$ {\color{darkgreen} \blacksquare}$ (\%) & 0.24 & 0.26 & 0.33 & 0.32 & 0.41 & 0.42 & 0.39  \\ 
	\hline
% 		${\color{orange} \blacksquare}
% 		-{\color{black} \blacksquare}$ (\%) 
% 		& 0.33 & 0.86 & 0.13 & 1.0 & 0.2 & 0.17 & 0.49  \\ 
% 		\hline
% 		${\color{darkgreen} \blacksquare}
% 		-{\color{black} \blacksquare}$ (\%)
% 		 & 0.11 & 0.08 & 0.09 & 0.11 & 0.1 & 0.15 & 0.08  \\ 
% 		\hline
	$ \frac{ {\color{orange} \blacksquare}
	-{\color{darkgreen} \blacksquare} } {
	{\color{orange} \blacksquare}
	-{\color{black} \blacksquare}
	} $ (\%) & 67.5 & 90.9 & 32.2 & 89.2 & 49.6 & 14.4 & 84.2  \\ 
	\hline
	\end{tabular}
	\caption{
	  Measured gate errors and the crosstalk error reduction ratio in each case corresponding to that in Fig.~\ref{fig:f4}b.
	}
\label{tab:t1}
}
\end{table}

To evaluate the performance of our method in the context of
	a more crowded system, we perform RB to an array of 
	7 qubits (Q1--Q7) on the same processor 
as depicted in Fig.~\ref{fig:f4}a (Q2 and Q4 were used in the previous demonstration).
The qubit frequencies are interleaved between 
	a lower-frequency ($\sim$5.6~GHz) 
	and higher-frequency ($\sim$6.2~GHz) band.
Since the neighboring qubits are largely detuned, 
	even though their crosstalk amplitude is relatively stronger, 
	the AC stark shift is still considerably smaller 
	than that between qubits in the same frequency band.
Therefore, we calibrate the crosstalk parameters 
	for each pair of qubits in the same band only.

We performed simultaneous RB on all 7 qubits without compensation and compare the extracted error per gate (EPG) with that measured separately (orange versus black bars in Fig.~\ref{fig:f4}b).
All qubits show higher EPG as 
	a result of crosstalk,
	for some qubits such as Q2 and Q4, the degradation is particularly significant,
	e.g., 6-fold EPG increase for Q4.
The observation can be explained by the AC stark shift matrix 
	(Fig.~\ref{fig:f4}c) calculated from the measured 
	crosstalk parameters and $\Omega=|r| \Omega_0$,
	where $\Omega_0=33$~MHz corresponding to the maximum Rabi frequency in a 30ns-long 
	cosine-shaped pulse used in the experiment. 
From the matrix, we can easily identify that the AC stark 
	effect induced by Q6 on Q2 and Q4 is much stronger 
	because of the adjacency in their frequencies.

Finally, we performed the same simultaneous RB 
	with crosstalk compensation turned on 
	(green bars in Fig.~\ref{fig:f4}b). 
All qubits show clear improvement 
	compared to the case without compensation 
	(green bars versus orange bars). 
In particular, most of the crosstalk error 
	in Q1, Q2, Q4 and Q7 are removed, 
	e.g., about 90\% for Q2 and Q4, 
	confirming the effectiveness of our method. 
The strongest crosstalk effect in these qubits 
	-- the maximum $\delta$ in each column of the matrix 
	-- comes from control qubits in the same frequency band 
	where the compensation is implemented.
Note that Q6 shows little improvement from compensation (only 14\% crosstalk error reduction), 
	even if the crosstalk effect from Q7 is still significant. 
	This is because Q6 and Q7 are in different frequency bands 
	and we did not implement inter-band compensation.
In addition to the microwave crosstalk, 
	the remaining errors (green bars versus black bars) 
	may come from quantum-mechanical crosstalk effect 
	such as the spectator effect
	\cite{cai2021impact,zhao2021quantum} 
	as the change in the state of neighboring qubits 
	can also cause a change in the frequency of the target qubit. 
n our device, there is on average a residual ZZ coupling strength of about 50~kHz. In addition, we note no performance degradation over time for the classical signal compensation, and the calibrated parameters are unaffected between different cool-downs as long as the wiring is unchanged.

To summarize, we successfully implement a method based on the AC Stark effect to calibrate the microwave signal crosstalk ubiquitously seen in superconducting quantum processors.
The method is an important addition to the calibration toolbox, especially suitable for qubits whose frequencies are non-tunable or subject to limited tunability.

We show that by crosstalk compensation we are able to remove the majority of crosstalk errors during simultaneous single-qubit gate operations.
We also find that such microwave crosstalk phenomenon can be highly non-local, affecting far-neighbor qubits when their frequencies are close. In a frequency-crowded system, it is crucial to fully calibrate such crosstalk for global high performance of the quantum processors.

\textbf{Acknowledgement:} 
This work was supported by 
the Key-Area Research and Development Program of Guang-Dong Province (Grant No. 2018B030326001), 
the National Natural Science Foundation of China (U1801661), 
the Guangdong Innovative and Entrepreneurial Research Team Program (2016ZT06D348), 
the Guangdong Provincial Key Laboratory (Grant No.2019B121203002), 
the Natural Science Foundation of Guangdong Province (2017B030308003), 
the Science, Technology and Innovation Commission of Shenzhen Municipality (KYTDPT20181011104202253), 
the Shenzhen-Hong Kong Cooperation Zone for Technology and Innovation (HZQB-KCZYB-2020050),
and the NSF of Beijing (Grant No. Z190012).
X.T. and Y.Y. acknowledge support from NSFC (Grant No. 61521001, No. 12074179 and No. 11890704) and NSF of Jiangsu Province (Grant No. 2021102024).

\textbf{Data availability:}
The data that support the plots within this paper and other findings of this study are available from the corresponding authors upon reasonable request.

\bibliography{apssamp}% Produces the bibliography via BibTeX.

\end{document}